\documentclass{jps-cp}
\usepackage{txfonts} 

\title{Spin Physics Detector at NICA}

\author{V.P.~\textsc{Ladygin}$^{1}$ on behalf of the SPD Collaboration}

\inst{$^{1}$Joint Institute for Nuclear Research, Dubna, Russian Federation}

\email{vladygin@jinr.ru}

\abst{The Spin Physics Detector  is one of two large setups at the
NICA collider under construction at JINR (Dubna). The ultimate goal of
the studies at SPD is the measurement of different spin observables in
polarized proton-proton and deuteron-deuteron  
collisions sensitive to the polarized gluon structure of the nucleon
at the luminosity up to 10$^{32}$~ $\mathrm{cm}^{-2}\cdot \mathrm{s}^{-1}$ and $\sqrt{\mathrm{s}}$ up to 27 GeV.
SPD will consist of the superconducting magnetic system, silicon
tracker, straw mini-drift
tubes tracker, time-of-flight system, electromagnetic ''shashlyk''- type
calorimeter, muon (range) and local-polarimetry systems. The high
performance free-streaming DAQ system will be able to operate at the
collision rate up to 4~MHz.
}

\kword{polarized beams, spin effects, nucleon gluon structure} 

\begin{document}
\maketitle

\section{Introduction}

The Nuclotron Based Ion Collider fAcility (NICA) devoted to relativistic nuclear, hadron and applied physics is under construction at the 
Joint Institute for Nuclear Research in Dubna, Russia \cite{nica}. 
NICA will accelerate different ions from proton to $\mathrm{Au}$  at the energies $\sqrt{\mathrm{s_{NN}}}$ up to 27 GeV and 11 GeV, respectively.   The beams
will collide in two interaction points, where the $\sim 4\pi$ detectors will be installed. The Multi-Purpose Detector (MPD) will be focused on the study
of the equation of the nuclear matter state at high nuclear densities and
moderate temperatures. The Spin Physics Detector (SPD) will be dedicated mostly to the physics with polarized ion beams in the  transition region between the perturbative and non- perturbative QCD.

The major goal of SPD \cite{SPD_cdr} is to study the spin
structure of the proton and deuteron and the other spin related phenomena with polarized proton and deuteron beams 
at the luminosity up to 10$^{32}$~ $\mathrm{cm}^{-2}\cdot \mathrm{s}^{-1}$ and at the collision energy $\sqrt{\mathrm{s}}$ up to 27 GeV. 
The kinematical region accessible at  SPD  covers the
transition region from non-perturbative to perturbative
QCD. The symmetries of the strong interaction, the
properties of the QCD vacuum, basic properties of particles as mass and spin will be studied through different processes. 
The kinematical range together with the high performances expected
from a modern $\sim 4\pi$ detector and availability of both polarized
proton and deuteron beams make the scientific program at SPD unique.

\section{Requirements to polarized beam facility}

The physics at SPD \cite{SPD_cdr} put serious requirements 
to the beams at NICA. 
The required energy ranges for $\mathrm{pp}$- and   $\mathrm{dd}$- collisions are 
$\sqrt{\mathrm{s}}$=  12$\sim$27~GeV (5$\sim$12.6 GeV kinetic energy) and 
$\sqrt{\mathrm{s}}$= 4$\sim$13~GeV (2$\sim$5.5 GeV/u kinetic energy), respectively.
The averaged luminosity for polarized $\mathrm{pp}$- 
collisions at $\sqrt{\mathrm{s}}$= 27~GeV to
be achieved is $\sim$10$^{32}$~ $\mathrm{cm}^{-2}\cdot \mathrm{s}^{-1}$ \cite{meshkov1}.
The beam lifetime and polarization degree required by SPD are a few hours and 
$\sim$70\%, respectively. 
Both longitudinal and transverse proton polarizations  at the SPD  
interaction point are needed. Asymmetric $\mathrm{pd}$- collision mode should be also available after NICA upgrade.

\begin{figure}[htbp]
\centering
{\includegraphics[width=10cm,clip]{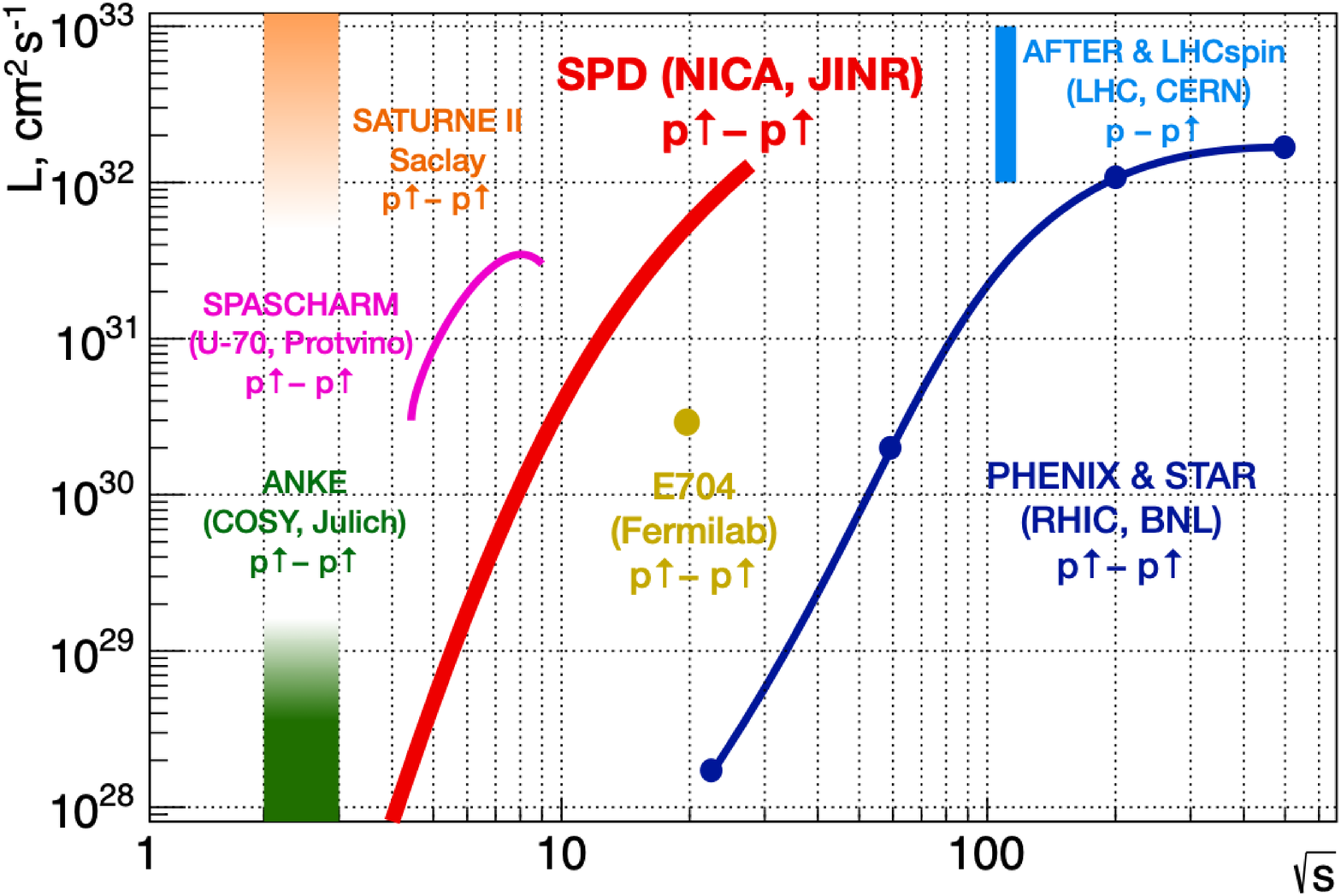} 
\caption{SPD at NICA and the other past, present, and future experiments with polarized protons.}}
\label{fig:fig1}
\end{figure}

The comparison of SPD with
other hadron world facilities is illustrated in Fig.\ref{fig:fig1}. 
The contribution to the world effort in understanding the strong interaction dynamics will be obtained by the measurements of
the variety of polarization observables that are accessible with colliding proton and deuteron beams  polarized along and transversal to the beam direction. The SPD experiment  at NICA \cite{SPD_cdr} will cover the kinematic
gap between the low- energy measurements at ANKE-COSY and SATURNE and the high- energy measurements at the Relativistic Heavy Ion Collider, as well as the planned fixed-target experiments
at the LHC.  The possibility for NICA to operate with polarized deuteron beams at such energies is unique.

The experimental results will be obtained by implementing the spin transparency
mode \cite{nica_ST}. Two solenoidal Siberian snakes will be installed in
the straight sections of the collider to control the polarization direction.
Only $\sim$12~T$\cdot$m   Siberian snakes in each ring will be used  at the first stage of SPD. This will provide  spin transparency
mode only up to $\sqrt{\mathrm{s}}\sim$6.7~GeV for $\mathrm{pp}$- collisions.  At higher 
energies spin transparency mode will be provided at the 
integer spin resonances \cite{nica_STk}. 
The detailed description of two schemes of the polarized proton beam formation  considered for the NICA is given in ref.\cite{syresin}. 

The new Source of Polarized Ions (SPI) \cite{NewPIS} has been  developed at JINR 
using part of the equipment of the IUCF Ion Source CIPIOS \cite{cipios}. 
It was  commissioned  and successfully used to provide polarized deuteron beam for the first priority  experiments at Nuclotron: ALPOM-2 \cite{alpom} and DSS \cite{dss1,dss2}. The typical values of the beam polarization were 
$\sim$65-75\% from the ideal values both for the vector and tensor components. 
The possibility to accelerate the polarized proton beam at Nuclotron at
500~MeV  has been also demonstrated. The measured value of the proton beam 
polarization was $\sim$40\% \cite{lad_pp}. 
Physics at SPD requires a significant increasing of the beam intensity as well as
the development of the spin orientation preserving system \cite{nuclotron_STk}.  
Therefore, the  facility operation in $\mathrm{pp}$- mode at $\sqrt{\mathrm{s}}$ = 27 GeV reaching
average luminosity of $\sim$10$^{32}$~ $\mathrm{cm}^{-2}\cdot \mathrm{s}^{-1}$  remains the first priority task for coming years.

\section{Scientific mission of SPD}

\subsection{Gluon probes} 

\begin{figure}[htbp]
\centering 
{\includegraphics[width=10cm,clip]{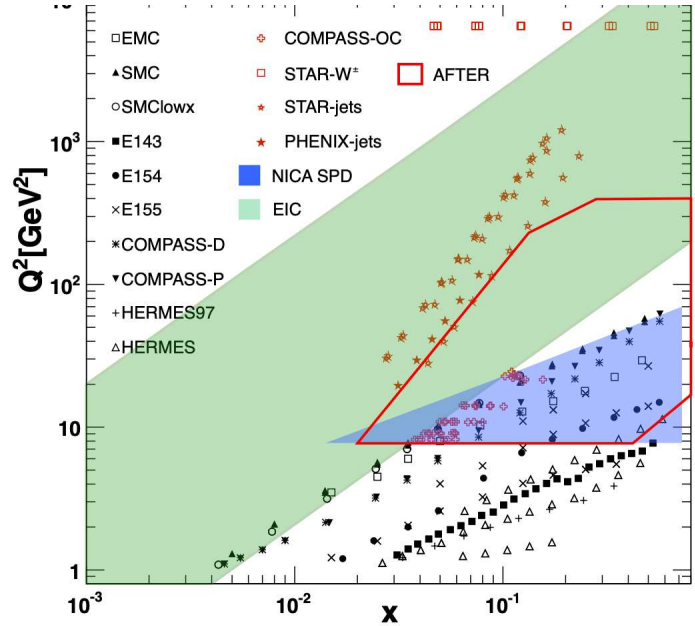} 
\caption{Kinematic coverage of the SPD in the charmonia, open charm, and prompt photon production
processes.}}
\label{fig:fig2}
\end{figure}

The polarized gluon content of proton and deuteron at intermediate and high values of the Bjorken $\mathrm{x}$ 
will be investigated using three complementary probes: inclusive production of charmonia, open charm, and prompt photons \cite{SPD_gluon}. 
The kinematic
phase-space in $\mathrm{x}$ and $\mathrm{Q^2}$ to be accessed by the SPD is compared to the corresponding ranges of previous, present and future experiments in Fig.\ref{fig:fig2}.

\begin{figure}[htbp]
\centering 
{\includegraphics[width=9cm,clip]{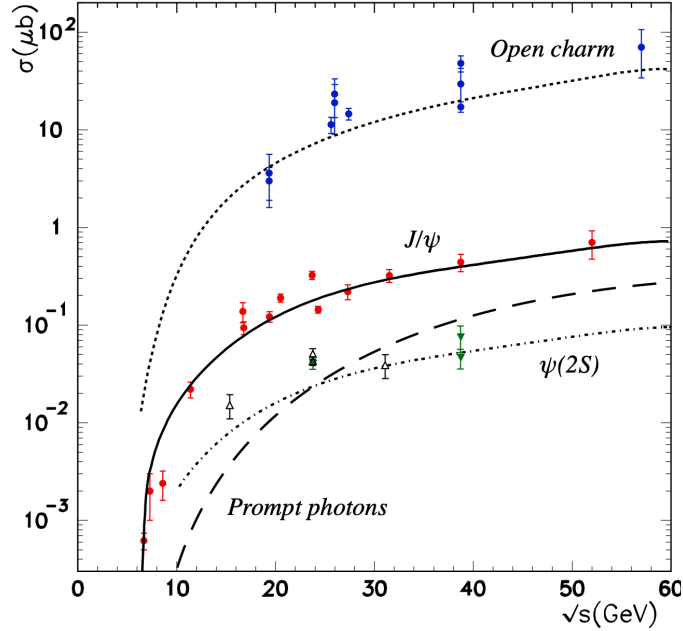} 
\caption{Cross-section for the processes of open charm,
$J/\psi$, $\psi(2S)$ and prompt photons ($\mathrm{p_T}\ge$ 3 GeV) production as a function of center-of-mass energy.}}
\label{fig:fig3}
\end{figure}

SPD is planned to operate as a universal facility for a comprehensive study of the unpolarized and polarized gluon content of the nucleon at large and 
moderate $x$. The experiment aims at providing
access to the gluon helicity, gluon Sivers, and Boer-Mulders parton distribution functions in the nucleon, as well as the
gluon transversity distribution and tensor PDFs in the deuteron, via the measurement of specific single and double spin asymmetries \cite{SPD_cdr}. 

The cross-section for the  open charm,  $J/\psi$ and 
prompt photons production processes is shown in Fig.\ref{fig:fig3}. 
The study of these three channels is complementary. 
The largest cross section is for open charm production. 
Since the D-meson is reconstructed from its decay channels,
$D^+\to\pi^+K^-\pi^+$  and  $D^0\to K^-\pi^+$, a precise secondary
vertex detection displaced by $\sim$100~$\mu$m from the interaction point
is required. 
The reaction $J/\psi\to \mu^+\mu^-$ gives a narrow signal 
over a background, with a relatively large cross section and branch-
ing ratio 6\%. A large fraction of $J/\psi$  is   produced
in the decays of heavier resonances, making the analysis
more complicated. 
Prompt photons are  carrying the information
of the gluon-quark or quark-antiquark interaction. 
However, they need to be disentangled from
a large background, particularly at low transverse 
momenta where photons from secondary mesons largely contribute.
This physics puts very high requirements on the detector.

The results expected to be obtained by SPD will play an
important role in the general understanding of the nucleon gluon content and will serve as a complementary input to the ongoing and planned studies at RHIC, and future measurements at the EIC (BNL)
and fixed-target facilities at the LHC (CERN). Simultaneous measurement of the same quantities using
different processes at the same experimental setup is of key importance for the minimization of possible
systematic effects.

\subsection{Physics at the first stage of the SPD}

SPD has an extensive physics program \cite{SPD_phase1} for the first stage of the NICA collider operation with
reduced luminosity and collision energy of the proton and ion beams, devoted to comprehensive
tests of the various phenomenological models in the non-perturbative and transition  region.

It includes such topics as the spin effects in $\mathrm{NN}$ elastic scattering, in exclusive reactions,  in hyperons inclusive production, multiquark correlations, di-quarks dynamics, dibaryon resonances, vector meson production, deuteron short-range spin structure, scaling properties of spin observables, diffractive and hard scattering,  physics of the light and intermediate nuclei collisions, hypernuclei production, dark matter searches  \cite{SPD_phase1}.

\section{Spin Physics Detector status}

\begin{figure}[htbp]
\centering 
{\includegraphics[width=14cm,clip]{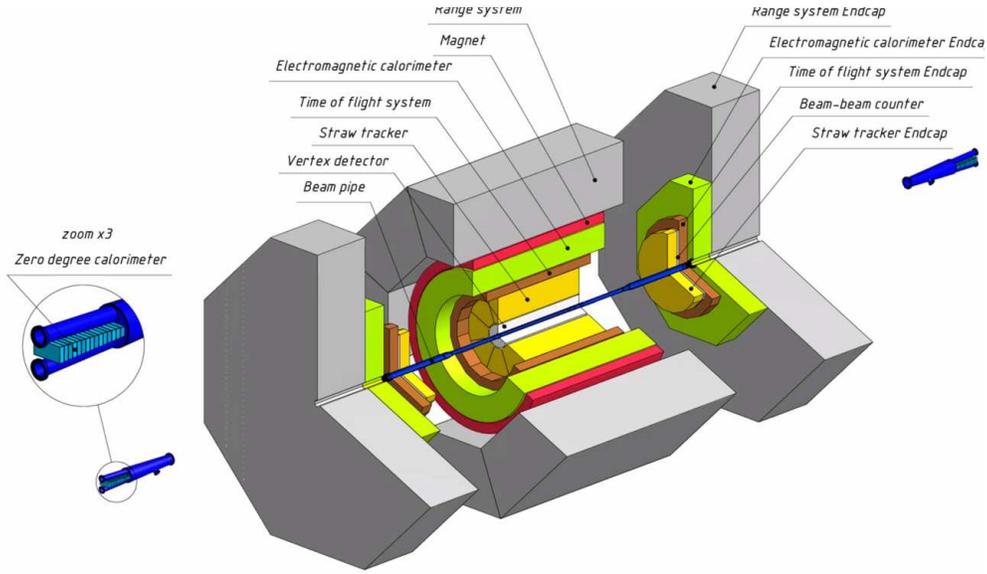} 
\caption{Layout of the SPD.}}
\label{fig:fig4}
\end{figure}

 The SPD   is being
designed as a universal 4$\pi$   experimental setup with advanced tracking and particle identification capabilities based on modern technologies.
It is shown schematically in Fig.\ref{fig:fig4}.
 The SPD will have a cylindrical
symmetry around the collider beam axis, set at the collision point, with longitudinal and transverse dimensions of $\sim$8~m and
6.5~m, respectively. The size of the
detector is limited by its weight of less than 1200 tons.

The detector will be embedded in a solenoidal magnetic field of $\sim$1~T at the axis. The main options of the magnetic system are the superconducting solenoid or a set of superconducting coils in a single cryostat that differs from the SPD CDR option \cite{SPD_cdr}.
The silicon vertex detector (VD) with a low material budget 
will provide the resolution for the
vertex position on the level of $\sim$100~$\mu$m needed for reconstruction of secondary vertices of $D$- meson decays.  VD will be placed 
as close as possible to the beryllium beam pipe 
(at distances of 5$\le$R$\le$25~cm). 
The use of Monolithic Active Pixel Sensors (MAPS) designed and produced for ALICE with the pixel size of 29~$\mu$m$\times$27~$\mu$m   improves the 
signal-to-background ratio of $D$- meson peak by a factor of 3. 
Micromegas technology is considered for the vertex detector at the first stage
of SPD operation. 
The tracking system (ST) based on the straw mini-drift  tubes and placed inside a solenoidal magnetic field should provide the transverse momentum resolution 
$\sigma_{p_T}$ /$p_T \approx$ 2\% for a particle momentum of 1 GeV/$c$. 
Information on the charged particles energy losses will be
used additionally for the identification of particles with the momenta 
$\le$0.7~GeV/$c$.

The time-of-flight system (PID) based on the mRPC
with a time resolution of about
60~ps will provide 3$\sigma$ $\pi$/K and K/p separation of up to about 
1.2 GeV/$c$ and 2.2 GeV/$c$, respectively.
Possible use of the aerogel-based Cherenkov detector in the endcaps could extend this range.
The sampling lead-scintillator electromagnetic calorimeter of the ''shashlyk''- type
(ECal)  
with a low
energy threshold of 50 MeV and an energy resolution of $\sim$5\%/$\sqrt{E}$ \cite{ECal} will be used for the detection of prompt photons and photons from particle decays, as well as for 
the identification of electrons.
To minimize multiple scattering and photon conversion effects for photons, the detector material will be kept to a minimum throughout the internal part of the detector.
 The range system (RS) \cite{RS} planned for muon identification is optimized 
for the $J/\psi\to \mu^+\mu^-$ decay. It can also act as a rough hadron calorimeter. 20 layers of $\mathrm{Fe}$ (3-6 cm) are interleaved with gaps for 
mini- drift tube detectors providing 2 coordinate readout. 
The total weight of RS is $\sim$800 tons,  it provides at least 4$\lambda_I$.
The pair of beam-beam counters (BBC) \cite{BBC}
will be responsible for the local
polarimetry by the detection of inclusive charged particles.  
Two zero-degree calorimeters placed at the distances $\sim$13~m from the 
detector center will be used as a luminosity monitor and for
ensuring the separation between neutrons and gammas. 

To minimize possible systematic effects, SPD will be equipped with a triggerless DAQ system. A high collision rate up to 4 MHz at the luminosity of $\sim$10$^{32}$~ $\mathrm{cm}^{-2}\cdot \mathrm{s}^{-1}$   poses a significant challenge to the DAQ, online monitoring, offline computing
system, and data processing software. We expect raw data stream of 20 GB/s (or 200 PB/year). 
The online filter will reduce data by the order of magnitude
up to $\sim$10 PB/year.

The detailed information on the performance of SPD can be found in 
ref. \cite{SPD_cdr}.

\section{Conclusions}

SPD  at NICA at JINR is  a
multipurpose $\sim 4\pi$ detector for the comprehensive 
QCD studies with polarized
proton and deuteron beams at $\sqrt{\mathrm{s}}$ up to 27 GeV.
SPD is a facility for the study of gluon content in
proton and deuteron at large $\mathrm{x}$ and the other spin related phenomena. 
It is an unique facility for polarized deuteron- deuteron collisions.

A strong tradition for polarized beams and targets exists at
JINR, where unique polarized proton, neutron
and deuteron beams are available in the GeV range.\\
  
This contribution is dedicated to the memory of Profs A.V.~Efremov and A.D.~Kovalenko.

\end{document}